\documentclass{article}

\usepackage{cite}
\usepackage{mathtools}
\usepackage{amssymb}
\usepackage{cleveref}
\usepackage{bm}
\usepackage{gensymb}
\usepackage{url}

\begin{document}
{\huge
\textbf\newline{Inverse scattering for reflection intensity phase microscopy}
}
\newline
\\
Alex Matlock\textsuperscript{1}*,
Anne Sentenac\textsuperscript{2},
Patrick C. Chaumet\textsuperscript{2},
Ji Yi\textsuperscript{3},
 and Lei Tian\textsuperscript{1}
\\
{\bf 1 Department of Electrical and Computer Engineering, Boston University, Boston, MA 02215, USA}
\\
{\bf 2 Institut Fresnel, Aix Marseille Univ., CNRS, Centrale Marseille, Marseille, France}
\\
{\bf 3 Department of Medicine, Boston University School of Medicine, Boston, MA 02215, USA}
\\
*amatlock@bu.edu


\section{Abstract}
Reflection phase imaging provides label-free, high-resolution characterization of biological samples, typically using interferometric-based techniques.
Here, we investigate reflection phase microscopy from {\it intensity}-only measurements under diverse illumination.
We evaluate the forward and inverse scattering model based on the first Born approximation for imaging scattering objects above a glass slide.
Under this design, the measured field combines {\it linear} forward-scattering and height-dependent {\it nonlinear} back-scattering from the object that complicates object phase recovery.
Using only the forward-scattering, we derive a linear inverse scattering model and evaluate this model's validity range in simulation and experiment using a standard reflection microscope modified with a programmable light source.
Our method provides enhanced contrast of thin, weakly scattering samples that complement transmission techniques.
This model provides a promising development for creating simplified intensity-based reflection quantitative phase imaging systems easily adoptable for biological research.

\section{Introduction}
Optical scatter-based imaging modalities have become vital as rapid, label-free research tools for characterizing biological morphology~\cite{boustany2010microscopic}. 
In particular, {\it reflection}-based systems are important because they: 1) measure backscattered fields carrying high frequency content sensitive to fine details in the object's axial structures~\cite{Jin.etal2017,boustany2010microscopic,beaurepaire1998full,cherkezyan2013interferometric,yi2015fractal,wang2010nanoscale,bista2012nuclear,Ortega-Arroyo:2012aa,dubois2002high,dubois2004ultrahigh} and 2) can be applied to {\it{ex vivo}} samples~\cite{wang2010nanoscale, boustany2010microscopic,subramanian2008optical,cherkezyan2013interferometric,yi2015fractal,bista2012nuclear,Daaboul:2010aa,Daaboul:2016aa,Avci:2017aa,Ortega-Arroyo:2012aa} and thick tissues {\it{in vivo}}~\cite{beaurepaire1998full,Kang2015, matthews2014deep, Huang1991,Ford2012, boustany2010microscopic, kandel2019epi}.
Recent modalities have expanded reflection systems to quantitative phase imaging (QPI) for recovering the object's refractive index (RI)~\cite{choma2005spectral,sarunic2006full, yamauchi2008low, edwards2014epi, wang2010nanoscale,yaqoob2011single, Zhang:2016aa,choi2014dynamic,choi2018reflection}.
Existing approaches often utilize interferometry~\cite{choi2014dynamic,yaqoob2011single,wang2010nanoscale,Ralston.etal2006,marks2007inverse,edwards2014epi,kandel2019epi,federici2015wide, uttam2015fourier} with successful high-resolution, high-sensitivity quantitative recovery of cellular and subcellular features~\cite{wang2010nanoscale,yamauchi2008low,edwards2014epi,yaqoob2011single,choi2014dynamic, kandel2019epi,yamauchi2011label,choi2018reflection}.
Interferometric techniques often require specialized optical setups that can be less accessible for certain biological applications.
In addition, QPI requires accurate scattering models for recovery, which vary significantly in reflection with the imaging modality design~\cite{elfouhaily2004critical,Kim.etal2015,zhou2017modeling,uttam2015fourier,marks2007inverse} and desired application~\cite{elfouhaily2004critical,Kim.etal2015,hu2017physical,avci2016physical,guerin2004second,sheppard1998imaging}. 
Specifically, the presence of boundaries or structures near the object can generate additional scattering requiring complex models~\cite{Zhang:2016aa, guerin2004second, mudry2010mirror, avci2016physical, simon2019versatile} or can result in transmission-like imaging conditions~\cite{Ford2012, ledwig2018dual,ledwig2019epi,laforest2019transscleral}.
These constraints suggest that QPI in reflection with standard microscope designs and computationally efficient, easily implementable inverse scattering models would be highly advantageous for biological research.
Recently, {\it intensity-only} techniques using diverse illumination paired with inverse scattering models have achieved great success for QPI in transmission and are easily built into standard microscopes~\cite{Tian2014, Tian2015a, Tian2015b, Tian.Waller2015, Chen2016, ling2018high, li2019high, horstmeyer2016diffraction, Matlock_2019}.
Here, we explored such an intensity-only reflection QPI approach using diverse illumination of a sample fixed on a glass slide.
We developed linear scattering models, built an optical setup, and evaluated our model in simulation and experiment.
We show linear models adequately recover thin objects at nanometer length-scales but become inaccurate from height-dependent nonlinear phase behavior in the measured back-scattered field.

\begin{figure}[t]
\centering
\includegraphics[width =0.86 \textwidth]{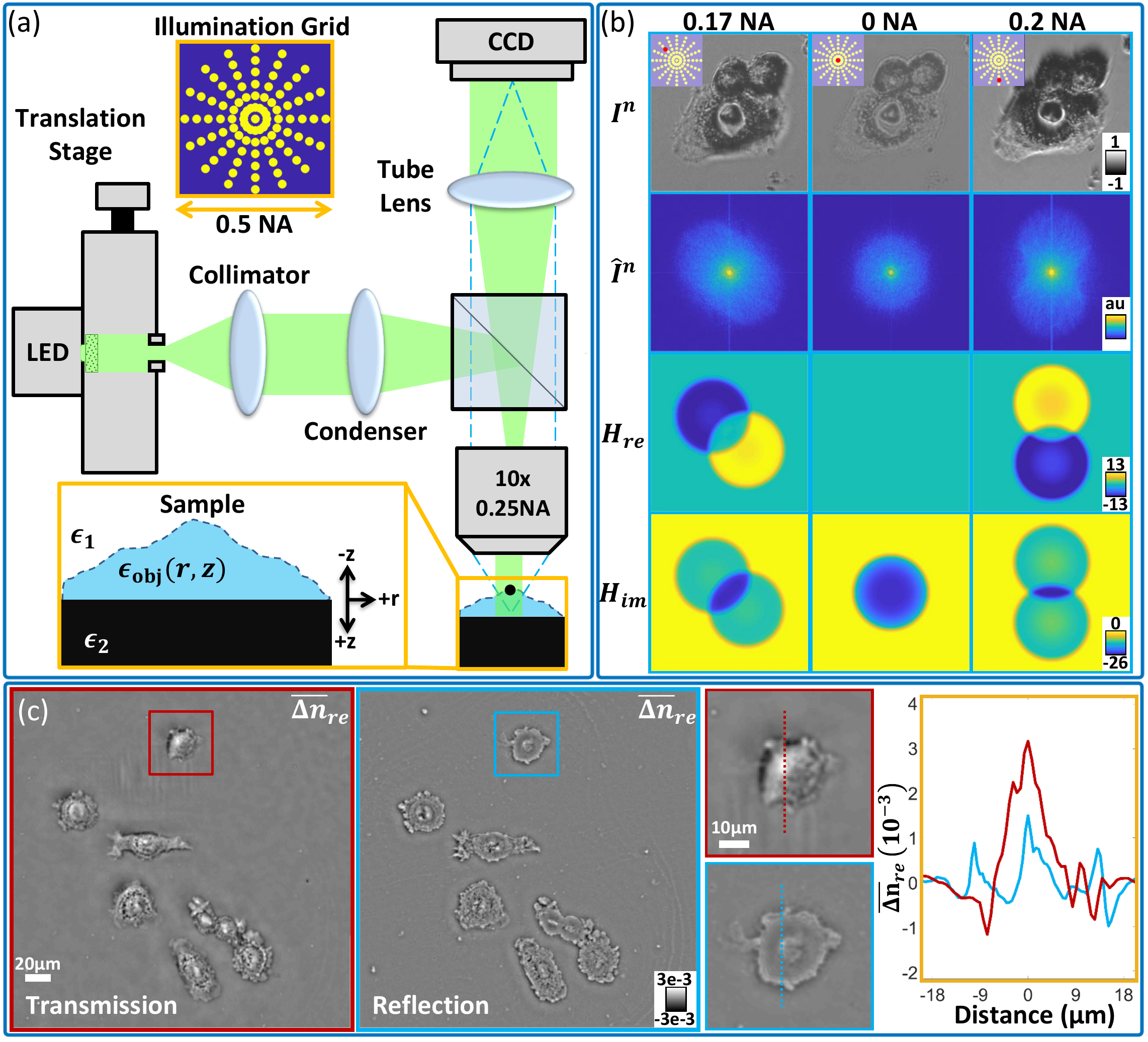}
\caption{(a) Reflection intensity phase microscope design with illumination grid and imaging geometry. A scannable LED in a conjugate plane to the objective's back focal plane enables programmable oblique illumination up to 0.25NA.
			(b) Normalized reflection images, Fourier coverage, and model transfer functions for illuminations at 0.17, 0, and 0.2NA. The phase transfer function show asymmetric behavior at oblique illumination and cancellation for on-axis illumination.
		     (c) The average real refractive index (RI) contrast reconstructions from transmission Intensity Diffraction Tomography~\cite{ling2018high} (Red) and our reflection system (Blue). Transmission better recovers large nuclear structures while reflection captures thin membrane features.
}
\label{F1}
\end{figure}

Intensity-only techniques recover the complex field information from ``phaseless'' measurements~\cite{Gbur:02ol, Gbur:02, ling2018high, soto2018optical, Tian.Waller2015, unger2019versatile}. 
They have been explored extensively in transmission~\cite{anastasio2004relationship, Gbur:02ol, Gbur:02,furhapter2005spiral,ling2018high,Tian.Waller2015} and rely on encoding phase information into intensity using defocus~\cite{Gbur:02ol, Gbur:02, streibl1984phase, huang2009image}, oblique illumination~\cite{ling2018high, Zheng2013, Tian.Waller2015,soto2018optical, horstmeyer2016diffraction, Matlock_2019}, and pupil engineering~\cite{zernike1942phase,furhapter2005spiral,papagiakoumou2010scanless}.
Here, we use oblique illumination combined with the system's pupil function to create asymmetric transfer functions (Fig.~\ref{F1}(b)) that render phase information visible~\cite{Tian2015a,Zheng2013,Tian2015b,Mehta2009,Hamilton1984a,Ford2012,ledwig2018dual,ledwig2019epi}.
This method enlarges the overall Fourier coverage from synthetic aperture principles~\cite{Zheng2013, Tian2015b}, improves the final reconstruction's resolution to the incoherent limit, and is easily implemented using a programmable light source in a standard microscope.
Using a system design (Fig.~\ref{F1}(a)) similar to the Fourier ptychography platforms of~\cite{Pacheco2016,guo2016Fourier,Pacheco2015}, we investigate computationally efficient {\it linear} inverse scattering models for directly recovering quantitative phase information in reflection. 


Linear reflection models have been explored previously for applications including metrology~\cite{elfouhaily2004critical} and quantitative tissue imaging~\cite{hu2017physical,Ralston.etal2006, zhou2017modeling, elfouhaily2004critical}.
Classical reflection models such as the Kirchhoff approximation are easily implemented and have found utility in metrology for confocal microscopy~\cite{wombell1991reconstruction,desanto1985exact,beckmann1987scattering,sheppard1993imaging}. 
This model is less applicable to biomedical imaging, where material inhomogeneities, optical roughness, and volumetric scattering of biological structures invalidate the approximation's underlying assumptions.
Furthermore, this model relies on direct field detection for quantitative recovery and requires interferometry-based designs or point-scanning the sample where phase from the object's height is unambiguous~\cite{wombell1991reconstruction,sheppard1993imaging}.
We instead evaluate a {\it volumetric} model using the first Born approximation for describing light scattered from an inhomogeneous object of variable height and permittivity.
This approximation is used elsewhere in reflection~\cite{hu2017physical,Ralston.etal2006, zhou2017modeling,guerin2004second,elfouhaily2004critical}, but we consider a partially reflective boundary interface below the object that creates additional scattering to model a biological sample fixed on a glass slide.
This case is similar to~\cite{guerin2004second, mudry2010mirror, unger2019versatile,simon2019versatile} where vectorial inverse scattering models are evaluated for nanoscale structures phase recovery.
These approaches are successful and evaluate both interferometric and intensity-only techniques but consider only nano-scale objects with complex optical setups or computationally intensive phase recovery techniques.
To develop an efficient model suitable for commonly encountered biological conditions, we consider a scalar model with a scattering object above a semi-infinite, partially reflecting interface.

Here, we evaluate a linearized reflection phase model and imaging modality for performing phase imaging in reflection.
We show this imaging condition captures both forward and back-scattered fields from the object that are {\it linear} and {\it nonlinear} with object height, respectively.
Using only the forward-scattered field information, we derive a linear inverse scattering model and show this field is twice as sensitive to the object's phase than transmission.
This enhanced phase sensitivity allows the recovery of objects at nanometer length-scales but induces a rapid breakdown of the model's validity range for increasingly tall objects.
We validate this behavior using rigorous discrete dipole approximation simulations of the imaging condition and illustrate its effects in experiment on fixed HeLa cells.
This work presents promising developments for QPI in reflection using simplified intensity-only imaging modalities.

\begin{figure}[t]
\centering
\includegraphics[width =0.8 \textwidth]{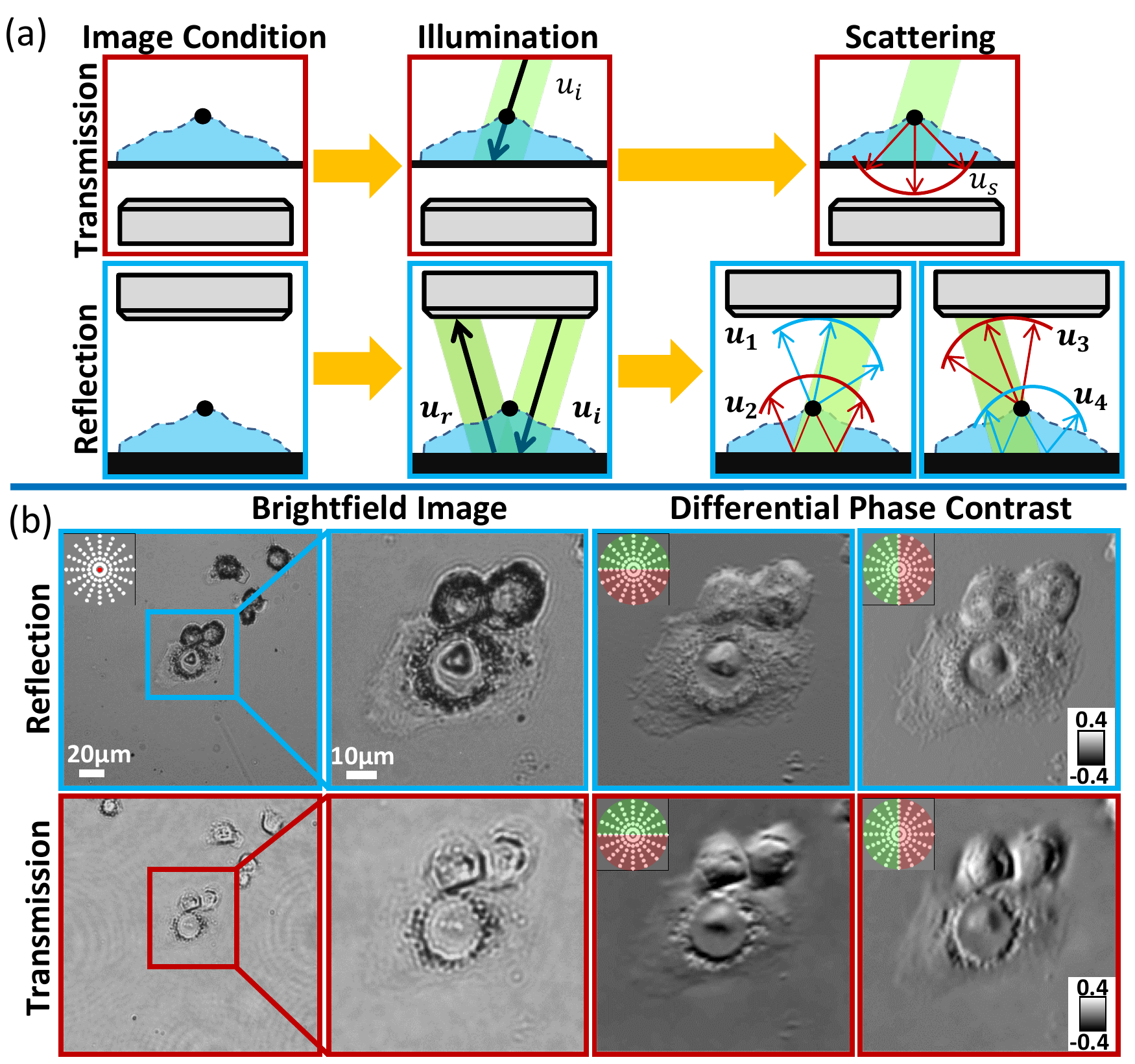}
\caption{(a) Illumination and expected scattering behavior under transmission and reflection geometries. 
		 (b) Comparison of on-axis brightfield and differential phase contrast (DPC) images of Henrietta Lacks (HeLa) cells in reflection and transmission. DPC images were generated from the difference of the images taken with the shown illuminations (Green - Red). The additional forward-scattering in reflection enhances thin cellular feature contrast.
}
\label{F2}
\end{figure}

\section{Theory}
We consider the imaging geometry in Fig.~\ref{F1}(a). 
An object of unknown permittivity and height is distributed between two homogeneous media with permittivities satisfying $\epsilon_1 < \epsilon_2$. 
We set the interface of these media as the $z=0$ plane with the object contained entirely in $\epsilon_1$ (i.e. $z\le0\mu$m). 
With this geometry, the total field must consider the reflections of both the object's scattered field and the illumination from the interface. 
Assuming quasi-monochromatic plane waves with central wavelength~$\lambda$ illuminate the object at arbitrary oblique angles up to the system's numerical aperture (NA),
the total illumination with this boundary is
\begin{equation}
	u_0(\bm{r},z|\bm{\nu_i}) =Ae^{j2\pi\bm{\nu_{i}}\cdot\bm{r}}\big(e^{j2\pi\eta(\bm{\nu_i})z} + R(\bm{\nu_{i}})e^{-j2\pi\eta(\bm{\nu_i})z}\big),
\end{equation}
	containing the incident plane wave $u_i$ and its reflection $u_{r}$ (Fig.~\ref{F2}(a)). 
Here, $\bm{r}$ and $z$ denote the lateral and axial spatial coordinates, respectively; 
	$\bm{\nu_i}$ and $\eta(\nu_i) = \sqrt{\lambda^{-2}-|\bm{\nu_i}|^2}$ are the illumination's lateral and axial spatial frequencies, respectively; 
	the amplitude $A = \sqrt{S(\bm{\nu_{i}})}P(-\bm{\nu_{i}})$; 
	$S$ is the incoherent primary source function; and $P$ is the pupil function. 
To calculate the reflected field, we use the average  TE and TM wave Fresnel coefficients
	$R(\bm{\nu_i})  = \frac{1}{2}[|R_{\mathrm{TE}}(\bm{\nu_i})| + |R_{\mathrm{TM}}(\bm{\nu_i})|]$
 to model the reflection amplitude of the unpolarized illumination from the boundary~\cite{sheppard1999fresnel}.

We consider a weakly scattering object with a {\it slowly varying} permittivity distribution~$\epsilon_{\mathrm{obj}}(\bm{r},z)$ characterized by the scattering potential $O(\bm{r},z) = {k_0^2}\Delta \epsilon(\bm{r},z)/{4\pi}$ with~$\Delta \epsilon (\bm{r},z) = \epsilon_{\mathrm{obj}}(\bm{r},z) - \epsilon_{1}$ and wavenumber $k_0 = 2\pi\lambda^{-1}$.  
Following the first Born approximation, the total field is
\begin{equation}
u_{\mathrm{\mathrm{tot}}}(\bm{r},z|\bm{\nu_i}) = u_{r}(\bm{r},z|\bm{\nu_i}) + \int_{-\infty}^{0} \iint_{-\infty}^{\infty} O(\bm{r'},z')u_0(\bm{r'},z'|\bm{\nu_i})G(\bm{r}-\bm{r'},z-z')d^2\bm{r'}dz',
\end{equation}
where the reflected illumination $u_{r}$ acts as a reference field and the second term describes the object's scattered field $u_s$ after illumination by the total incidence $u_0$.
Because of the boundary interface, we use the half-space Green's function $G$ in Weyl expansion form~\cite{guerin2004second} to account for both the forward and back-scattered fields (Fig.~\ref{F2}(a))
\begin{equation}
G(\bm{r}-\bm{r'},z-z') = j \iint_{-\infty}^{\infty} \frac{1}{\eta(\bm{\nu})} e^{j2\pi(\bm{\nu}\cdot(\bm{r}-\bm{r'}) - \eta(\bm{\nu}) z)}[e^{j2\pi\eta(\bm{\nu}) z'} + R(\bm{\nu})e^{-j2\pi\eta(\bm{\nu}) z'}]d^2\bm{\nu},
\end{equation}
where $\bm{\nu}$ and $\eta(\bm{\nu}) = \sqrt{\lambda^{-2} - |\bm{\nu}|^2}$ are the scattered field's lateral and axial spatial frequencies, respectively, and $R(\bm{\nu})$ is the scatter-angle ($\bm{\nu}$) dependent Fresnel coefficient. 
We constrain $z'$ over the object's height $h(\bm{r'})$ by $z' \in [-h(\bm{r'}),0]$ to obtain the scattered field 
\begin{align}
& u_s(\bm{r},z|\bm{\nu_i}) = j\frac{Ak_0^2}{4\pi} \iint_{-\infty}^{\infty} \iint_{-\infty}^{\infty} \int_{-h(\bm{r'})}^{0} \frac{1}{\eta(\bm{\nu})} e^{j2\pi(\bm{\nu}\cdot\bm{r} - \eta(\bm{\nu}) z )}\Big{[}e^{j2\pi \eta^{+}(\bm{\nu})z'} + R(\bm{\nu_{i}})e^{j2\pi \eta^{-}(\bm{\nu})z'}\nonumber \\
& + R(\bm{\nu})e^{-j2\pi \eta^{-}(\bm{\nu})z'} 
 +  R(\bm{\nu_{i}})R(\bm{\nu})e^{-j2\pi \eta^{+}(\bm{\nu})z'}\Big{]} \Delta \epsilon(\bm{r'},z')e^{-j2\pi \bm{\nu^{-}}\cdot \bm{r'}}dz'd^2\bm{r'}d^2\bm{\nu}, \label{usRaw}
\end{align}
where $\bm{\nu^{-}} = \bm{\nu} - \bm{\nu_i}$, $\eta^{+}(\bm{\nu}) = \eta(\bm{\nu}) + \eta(\bm{\nu_i})$, and $\eta^{-}(\bm{\nu}) = \eta(\bm{\nu}) - \eta(\bm{\nu_i})$. 
For simplicity, we assume the object's permittivity is uniform and constant over $[0, h'(\bm{r})]$.
Following axial integration and a Fourier transform of the field, we obtain the Fourier scattered field
\begin{equation}
	U_s(\bm{\nu},z|\bm{\nu_i})=  j\frac{Ak^2_0}{4\pi} \frac{e^{-j2\pi\eta(\bm{\nu}) z }}{\eta(\bm{\nu})} \iint_{-\infty}^{\infty} \Big{[}\mathrm{F}(\bm{\nu},\bm{r}'|\bm{\nu_i})+\mathrm{B}(\bm{\nu},\bm{r}'|\bm{\nu_i})\Big{]}e^{-j2\pi \bm{\nu^{-}}\cdot \bm{r'}}d^2\bm{r'}, \label{UsProb}
\end{equation}
where
\begin{align}
\mathrm{F}(\bm{\nu},\bm{r}'|\bm{\nu_i}) &= \Delta \epsilon(\bm{r}')h(\bm{r}')
\mathrm{sinc}(\pi \eta^-(\bm{\nu})h(\bm{r}'))[R(\bm{\nu_i})e^{-j\pi\eta^-(\bm{\nu})h(\bm{r}')} + R(\bm{\nu})e^{j\pi\eta^-(\bm{\nu})h(\bm{r}')}] \label{F} \\ 
\mathrm{B}(\bm{\nu},\bm{r}'|\bm{\nu_i}) &= \Delta \epsilon(\bm{r}')h(\bm{r}')
\mathrm{sinc}(\pi \eta^+(\bm{\nu})h(\bm{r}'))[e^{-j\pi\eta^+(\bm{\nu})h(\bm{r}')} + R(\bm{\nu})R(\bm{\nu_i})e^{j\pi\eta^+(\bm{\nu})h(\bm{r}')}] \label{B}
\end{align}
denote the {\it slowly} oscillating forward-scattering ($\mathrm{F}$) and {\it rapidly} oscillating back-scattering ($\mathrm{B}$) axial phase contributions.
$\mathrm{F}$ and $\mathrm{B}$ map to the lower and upper regions of the sample's Fourier space providing low and high-resolution information of the axial features, respectively~\cite{Wolf1969}.
This additional feature content enhances the captured phase information over transmission (Fig.~\ref{F2}), but the rapidly oscillating phase in $\mathrm{B}$ and non-elementary integration due to $h(\bm{r}')$ make quantitative phase recovery difficult and possibly nonlinear or ambiguous~\cite{hu2017physical}.
Thus, additional assumptions on the field behavior are required to maintain a simplified physical model.

\begin{figure}[t]
\centering
\includegraphics[width =1 \textwidth]{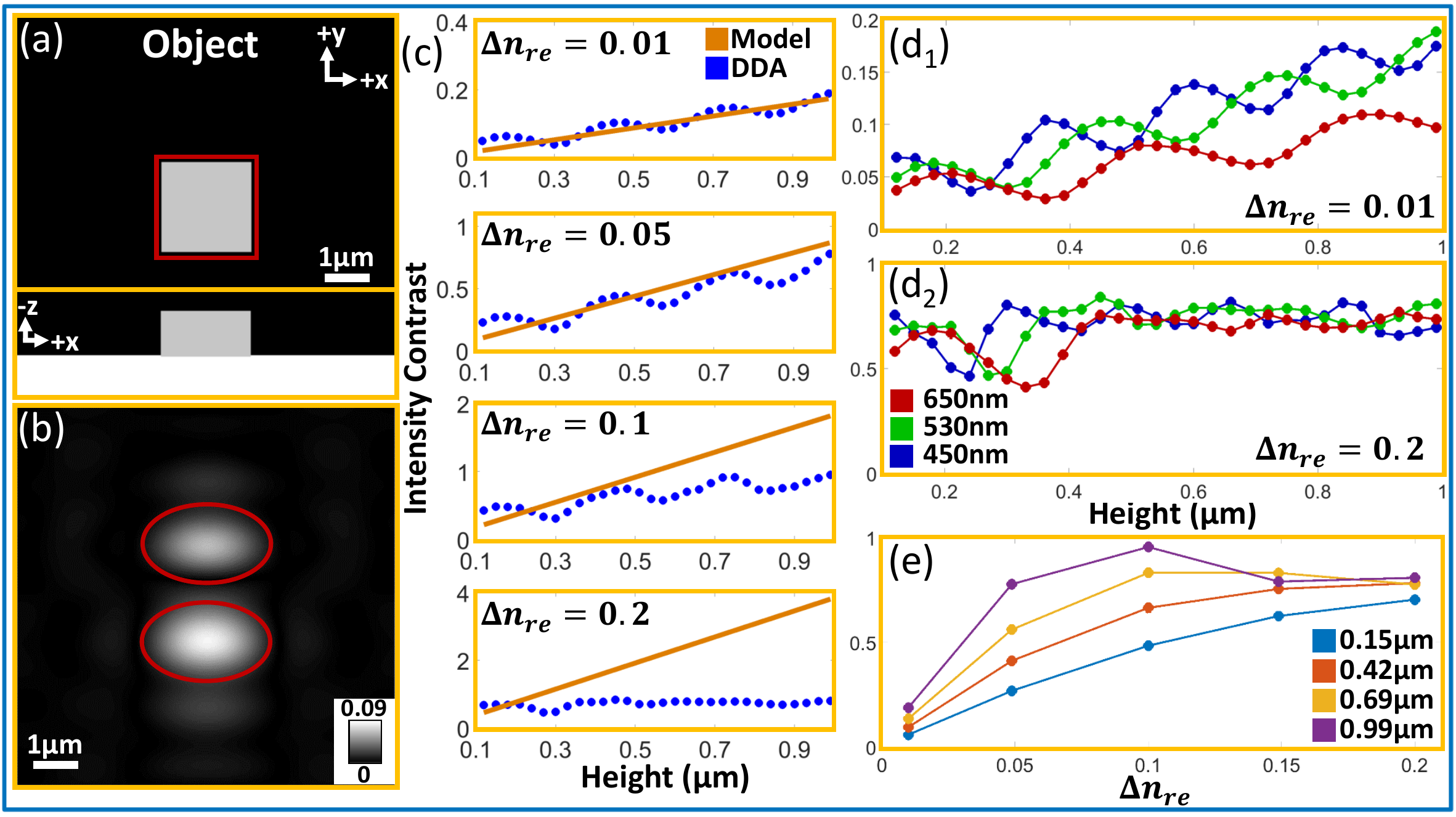}
\caption{(a) 3D Cuboid distribution above partially reflective surface from DDA simulations.
		 (b) Cuboid intensity contrast ($\Delta n_{\mathrm{re}}=0.01$, $h=210nm$, $\lambda=530nm$) with red ovals highlighting evaluated contrast region.
		 (c) Linear reflection model (Orange) and DDA simulation (blue) intensity contrast under 0.2NA illumination for objects with heights $0.12-1\mu m$ and increasing RI contrast for $\lambda=530nm$. The linear model adequately predicts the contrast at weak object permittivities but overestimates larger RI object contrast.
		 (d) Intensity contrast at fixed real RI ($\Delta n_{\mathrm{re}} = 0.01,0.2$ for d$_1$,$_2$ respectively) with increasing height across multiple wavelengths. The nonlinear term's period follows $\lambda/2$ until high RI contrast objects are evaluated.
		 (e) Intensity contrast highlighting linear trends for increasing RI contrast at fixed object heights.}
\label{F3}
\end{figure}

To determine appropriate assumptions, we use a Discrete Dipole Approximation (DDA)~\cite{draine1994discrete} model  from Marseilles Fresnel Institute~\cite{khadir2019quantitative, chaumetcode2019} to rigorously simulate our imaging condition from first principles.
We simulate $1.8 \times 1.8 \mu m^2$ cuboid objects (Fig.~\ref{F3}(a)) with varying heights ($0.12-1\mu m$) and real permittivity contrasts ($\Delta \epsilon=0.02-0.44$) on a 256$\times$256$\times$256 pixel grid with 30$nm$ sampling.
We convert the object's permittivity to refractive index (RI) contrast in our results, as it is more commonly used  in the QPI field.
The corresponding real RI contrast has a range of $\Delta n_{\mathrm{re}}=0.01-0.2$.

We generate simulated intensity images for each object using a 0.25NA, 10$\times$ magnification objective and oblique illuminations up to 0.2 NA.
This design and maximum illumination angle matched our experimental setup's maximum illumination as discussed in greater detail below.
Because the intensity encodes greater phase and scattering information from the object with oblique illumination (Fig.~\ref{F1}(b)), we evaluate the image contrast under 0.2 NA illumination for each simulated object (Fig.~\ref{F3}(b)).
We use the average of the image's full-width half-maximum (FWHM) as our contrast metric
\begin{equation}
	C = \overline{FWHM(|I^N|)},
\end{equation}
where $I^N = (I - \bar{I})/\bar{I}$ is the normalized, background-subtracted intensity image.
This approach captures the object's average scattering contrast without significant influence from the low-valued background or extreme, high-valued saturated pixels. 

The recovered intensity contrast values for various simulations are shown in Figure~\ref{F3}(c)-(e).
We evaluate the contrast across different object heights at fixed RI and wavelength $\lambda=530nm$ (Fig.~\ref{F3}(c)), across different imaging wavelengths (Fig~\ref{F3}(d)), and with increasing RI at fixed object heights and wavelength $\lambda=530nm$ (Fig.~\ref{F3}(e)).
For weakly scattering objects, we observe {\it linearly} increasing scattering contrast plus a {\it nonlinear} sinusoidal oscillation with respect to the object's height (Fig.~\ref{F3}(c) blue,(d$_1$)).
Across different imaging wavelengths, this height-dependent nonlinearity is preserved with an oscillation period matching $\lambda/2$ (Fig.~\ref{F3}(d$_1$)).
With increasing RI contrast, this behavior breaks down across all imaging wavelengths as the object becomes strongly scattering (Fig.~\ref{F3}(c),(d$_2$)).
Furthermore, the intensity contrast shows a linear relationship with increasing RI contrast at fixed heights until the object becomes strongly scattering (Fig.~\ref{F3}(e)).

These results agree with our Born-based derivation in Eq.~\eqref{UsProb} but show the difficulty of reflection QPI with linear models for intensity-only measurements.
The linear intensity trends in object height and RI can be attributed to the forward-scattered phase of Eq.~\eqref{F}.
This phase is inherently nonlinear, but its slowly varying nature is adequately approximated with linear functions as shown in transmission systems~\cite{ling2018high,Tian.Waller2015}.
This results in a directly proportional, linear relationship between the object's physical parameters and the field amplitude.
The oscillating nonlinear intensity contrast results from the backscattering phase of Eq.~\eqref{B}.
This field oscillates with a period of $\lambda/2$ as a function of object height, matching the behaviors observed in Fig.~\ref{F3}(c)-(d).
The inherent nonlinearity of the backscattering means that a linear model is insufficient to capture the full scattered field behavior for quantitative recovery in reflection.
We can therefore only recover the {\it forward-scattered} object features with a linear model for this imaging condition.

\subsection{Linear Scattering Model}
To evaluate whether a linear reflection model still provides relevant morphological information, we generate a forward model that ignores the backscattered field contribution of Eq.~\eqref{B}.
Specifically, we consider this entire backscattered term as an unrecoverable nonlinear error that reduces our linear model's validity.
Without this nonlinear term in Eq.~\eqref{UsProb}, we assume the forward-scattered field in Eq.~\eqref{F} will slowly accumulate phase through the object and can be linearized.
Following an application of Euler's formula and the definition of the ${\rm sinc}(\cdot)$ function, we obtain the reformulation of Eq.~\eqref{F}:
\begin{equation}
\mathrm{F}(\bm{\nu},\bm{r}'|\bm{\nu_i})=\frac{\Delta \epsilon(\bm{r}')}{j2\pi\eta^-(\bm{\nu})}\big{[}R(\bm{\nu_i})(1 - e^{-j2\pi\eta^-(\bm{\nu})h(\bm{r}')}) + R(\bm{\nu})(e^{j2\pi\eta^-(\bm{\nu})h(\bm{r}')}-1)\big{]}
\end{equation}
from which a Taylor expansion on the height-dependent exponential terms provides the field linearization: $\mathrm{F}(\bm{\nu},\bm{r}'|\bm{\nu_i})=\Delta \epsilon(\bm{r}')h(\bm{r}')[R(\bm{\nu_i}) + R(\bm{\nu})]$.
Solving Eq.~\eqref{UsProb} with this term, we obtain a linear Fourier scattered field
\begin{equation}
U_s(\bm{v},z|\bm{\nu_i}) = j\frac{Ak_0^2}{4\pi}\frac{e^{-j2\pi\eta(\bm{\nu}) z}}{\eta(\bm{\nu})}\mathcal{R}(\bm{\nu}|\bm{\nu_i})\widehat{\Phi}(\bm{\nu^{-}}), \label{us2D}
\end{equation}
where $\widehat{\Phi}(\bm{\nu^{-}}) = \mathcal{F}\{\Delta \epsilon (\bm{r})h(\bm{r})\}$ is the Fourier transform ($\mathcal{F}$) of the object-dependent phase introduced to the scattered field and is discussed in greater detail below, while the lateral frequency variable $\bm{\nu^{-}}$ describes the frequency shifted scattered field from oblique illumination providing enlarged Fourier coverage, akin to synthetic aperture. 
 $\mathcal{R}(\bm{\nu}|\bm{\nu_{i}}) = [R(\bm{\nu_{i}}) + R(\bm{\nu)}]$ is the modified Fresnel coefficient that accounts for the two forward-scattered field contributions.
The phase term $e^{-j2\pi\eta(\bm{\nu}) z}$ accounts for the additional phase induced by the sample defocus $z$.

With this field, we derive transfer functions (TF) for the image intensity to complete our linear model.
Assuming $L$ oblique illuminations, the intensity $I_l$ in the $l^{\mathrm{th}}$ image is the result of the total field $u_{\mathrm{tot}}$ filtered by the pupil function $P$ 
\begin{equation}
I_l(\bm{r},z|\bm{\nu_i}) = |\mathcal{F}^{-1}\{U_{\mathrm{tot},l}(\bm{\nu},z|\bm{\nu_i})P(\bm{\nu})\}|^{2}, \label{IntDef}
\end{equation} 
which contains four terms including the intensities of the reference field, scattered field, and their interference terms. 
With our first Born approximation, we neglect the weak intensity contribution from the scattered field's intensity and perform background subtraction to remove the reference intensity. 
The remaining interference terms describe a {\it linear} relation between the object permittivity and the measured intensity. 

We decompose the object's complex permittivity contrast $\Delta \epsilon(\bm{r},z) = \Delta \epsilon_{\mathrm{re}}(\bm{r},z) + j\Delta \epsilon_{\mathrm{im}}(\bm{r},z)$ into real and imaginary components and solve for them separately, following~\cite{Tian2015a,ling2018high}. 
These terms carry different physical meanings and are considered decoupled and separable during reconstruction.
We obtain the Fourier scattered field with respect to our TFs using normalized, background-subtracted intensity images $I^N_l$
\begin{equation}
\widehat{I}^N_l(\bm{\nu},z|\bm{\nu_i}) =
	C\big{[}
		\mathcal{H}_{\mathrm{re}}(\bm{\nu},z|\bm{\nu_{i}})\widehat{\Phi}_{\mathrm{re}}(\bm{\nu}) + \mathcal{H}_{\mathrm{im}}(\bm{\nu},z|\bm{\nu_{i}})\widehat{\Phi}_{\mathrm{im}}(\bm{\nu})
	\big{]} \label{Fwd},
\end{equation}
where $C = -(|A|^2R(\bm{\nu_{i}})k_0^2)/4\pi$ is a constant coefficient. 
The real and imaginary TFs are
\begin{subequations}
\begin{align}
\mathcal{H}_{\mathrm{\mathrm{im}},l}(\bm{\nu},z|\bm{\nu_{i}}) &= 
	P(\bm{\nu_{i}})P^*(\bm{\nu^{-}})D(\bm{\nu}^{-},z)\mathcal{R}(\bm{\nu^{-}}|\bm{\nu_i}) 
	+P^*(\bm{\nu_{i}})P(\bm{\nu^{+}})D^*(\bm{\nu}^{+},z)\mathcal{R}(\bm{\nu^{+}}|\bm{\nu_i}), \\
\mathcal{H}_{\mathrm{re},l}(\bm{\nu},z|\bm{\nu_{i}}) &= 
		j\{P(\bm{\nu_{i}})P^*(\bm{\nu^{-}})D(\bm{\nu}^{-},z)\mathcal{R}(\bm{\nu^{-}}|\bm{\nu_i}) 
		- P^*(\bm{\nu_{i}})P(\bm{\nu^{+}})D^*(\bm{\nu}^{+},z)\mathcal{R}(\bm{\nu^{+}}|\bm{\nu_i})\}. \label{H2D}
\end{align}
\end{subequations}
where $D(\bm{\nu},z) = e^{j2\pi\eta^{-}(\bm{\nu})z}/\eta(\bm{\nu})$ is the objective's defocus with an obliquity factor.

This linear model predicts a contrast enhancement in reflection over transmission.
The model recovers the sample's phase $\Phi_{\mathrm{re}}(\bm{r}) = \Delta \epsilon_{\mathrm{re}}(\bm{r}) h(\bm{r})$, which is related to the sample's optical path length:~$\Delta \epsilon_{\mathrm{re}}(\bm{r})h(\bm{r})~\approx~2\Delta n(\bm{r})n_1(\bm{r})h(\bm{r})$ where ($n_1 = \sqrt{\epsilon_1}$, $n_{\mathrm{re}} = \sqrt{\epsilon_{\mathrm{re}}}$, and $\Delta n = n_{\mathrm{re}} - n_1$).
This result resembles that of~\cite{sheppard1993imaging} and shows the phase through the object is doubled compared to transmission.
Thin features typically unobservable in transmission will thus provide better contrast in reflection from accumulating additional phase in the forward-scattered field.
This is observed qualitatively in the differential phase contrast (DPC)~\cite{Tian2015a} images of Fig.~\ref{F2}(b), where the cellular membrane and filopodial structures are more apparent in reflection.
With more rapid phase accumulation, however, the linearity range of the forward-scattered phase is halved.
We thus expect a reduced range of object heights recoverable with this linearized model. 

In addition, this model includes synthetic aperture behavior from oblique illumination with two shifted pupil functions, $P$ and its complex conjugate $P^*$, centered at $\bm{\nu^{+}} = \bm{\nu} + \bm{\nu_{i}}$, and $\bm{\nu^{-}} = \bm{\nu} - \bm{\nu_i}$.
These functions exhibit frequency shifts in opposite directions based on the illumination spatial frequency $\pm \bm{\nu_i}$.
As shown in the intensity image $I^n$ and the normalized intensity spectra $\hat{I}^n$ of Fig.~\ref{F1}(b), the use of oblique (0.2 NA) versus on-axis (0 NA) illumination captures higher resolution information by enhancing the recovered object's bandwidth.
This enhancement follows synthetic aperture principles~\cite{ling2018high} and can be extended by increasing the illumination angle until the incident field exceeds the objective NA.
At illuminations matching the objective NA, the pupil function shifts by the objective NA to achieve a maximum resolution of $\lambda/2NA$ matching the incoherent resolution limit.
In practice, hardware limitations prevent these high angles and less oblique illuminations are acquired.
We encounter this limitation here and use lower angle illumination as discussed below.

The TFs also exhibit different symmetries as previously observed in transmission~\cite{Tian2015a,ling2018high}. 
The real and imaginary TFs are asymmetric and symmetric, respectively.
As the real TF recovers the object's {\it phase}, the TF's asymmetry provides increasingly better phase contrast and recovery with larger oblique illuminations (Fig.~\ref{F1}(b))~\cite{ling2018high, Tian2015a}.
The imaginary TF is symmetric and recovers object features generating a loss of energy to the total illumination.
Because of the reflection imaging condition, this TF recovers both the object's {\it absorbing} features and the object's {\it reflectivity}.
This behavior is shown experimentally in Fig.~\ref{F5} where the cell's scattering structures are present in both the real and imaginary reconstructions.

Finally, the phase of $D(\bm{\nu},z)$ is best understood under paraxial conditions with oblique illumination:~$\eta^{-}(\bm{\nu}^{+})\approx-\lambda(\bm{\nu_i}\cdot\bm{\nu}+|\bm{\nu}|^2/2)$.
This term provides a linear geometric shift akin to lightfield~\cite{Tian2014a} and Fresnel diffraction, respectively.
These factors enable the post-correction of focusing errors during object reconstruction and have similar form for $\eta^{-}(\bm{\nu}^{-})$.

To evaluate this forward model, we generate the expected intensity contrast using Eq.~\eqref{Fwd} on the same objects and imaging conditions as the DDA simulation in Fig.~\ref{F3}(c)(Orange).
Our model adequately estimates the intensity contrast's linear component for objects with weak phase (Fig.~\ref{F3}(c)) but quickly overestimates more strongly scattering objects.
Since these overestimated objects are typically considered weakly scattering in transmission, our model's failure in this range highlights its reduced validity range from the enhanced phase sensitivity.
When reconstructing the object with this reflection model, we will thus recover thin structures and underestimate taller objects.
We confirm this in simulation and experiment in  Section~\ref{sec:results} using the reconstruction method described below.

\subsection{Object Reconstruction}
For recovering the object phase, we consider all $L$ measurements and implement Tikhonov deconvolution for recovering the complex object by solving
\begin{align}
\underset{\widehat{\Phi}_{\mathrm{re}},\widehat{\Phi}_{\mathrm{\mathrm{im}}}}{\mathrm{min}} 
	\sum_{l=1}^{L} 
		\big{|}\big{|}\widehat{I}^N_l - 
		(\mathcal{H}_{\mathrm{im},l}\widehat{\Phi}_{\mathrm{\mathrm{im}}} 
		+ \mathcal{H}_{\mathrm{re},l}\widehat{\Phi}_{\mathrm{re}})\big{|}\big{|}^2_2
	+\tau_{\mathrm{\mathrm{im}}}\big{|}\big{|}\widehat{\Phi}_{\mathrm{\mathrm{im}}}\big{|}\big{|}^2_2 
	+ \tau_{\mathrm{re}}\big{|}\big{|}\widehat{\Phi}_{\mathrm{re}}\big{|}\big{|}^2_2,
\end{align}
where $\tau_{\mathrm{im}}$ and $\tau_{\mathrm{re}}$ are regularization parameters. 
We obtain the following closed-form solution
\begin{subequations}
\begin{align}
&\hspace{-10pt} {\Phi}_{\mathrm{re}}(\bm{r}) = 
	\mathcal{F}^{-1}\Bigg{\{}
				\frac{1}{T}
				\bigg{[}
					\Big{(}\sum_{l=1}^L \big{|}\mathcal{H}_{\mathrm{im},l}\big{|}^2 + \tau_{\mathrm{\mathrm{im}}}\Big{)}
					\Big{(}\sum_{l=1}^L \mathcal{H}_{\mathrm{re},l}^*\widehat{I}^N_l\Big{)}
				  - \Big{(}\sum_{l=1}^L \mathcal{H}_{\mathrm{re},l}^* \mathcal{H}_{\mathrm{im},l}\Big{)}
					\Big{(}\sum_{l=1}^L \mathcal{H}_{\mathrm{im},l}^* \widehat{I}^N_l\Big{)}
				\bigg{]}
	\Bigg{\}} 
\label{oreal}   \\
&\hspace{-10pt} {\Phi}_{\mathrm{\mathrm{im}}}(\bm{r}) =
	\mathcal{F}^{-1}\Bigg{\{}
		\frac{1}{T}\bigg{[}
			\Big{(}\sum_{l=1}^L \big{|}\mathcal{H}_{\mathrm{re},l}\big{|}^2 
			+ \tau_{\mathrm{re}}\Big{)}
			\Big{(}\sum_{l=1}^L \mathcal{H}_{\mathrm{im},l}^*\widehat{I}^N_l\Big{)}
			 -\Big{(}\sum_{l=1}^L \mathcal{H}_{\mathrm{im},l}^* \mathcal{H}_{\mathrm{re},l}\Big{)}
			\Big{(}\sum_{l=1}^L \mathcal{H}_{\mathrm{re},l}^* \widehat{I}^N_l\Big{)}
		\bigg{]}
	\Bigg{\}} \label{oimag}
\end{align}
\end{subequations}
where 
$T = \big{(}\sum_{l=1}^L |\mathcal{H}_{\mathrm{re},l}|^2 + \tau_{\mathrm{re}}\big{)}
	\big{(}\sum_{l=1}^L |\mathcal{H}_{\mathrm{im},l}|^2 + \tau_{\mathrm{\mathrm{im}}}\big{)}
  - \big{(}\sum_{l=1}^L \mathcal{H}_{\mathrm{re},l} \mathcal{H}_{\mathrm{im},l}^*\big{)}
	\big{(}\sum_{l=1}^L \mathcal{H}_{\mathrm{re},l}^* \mathcal{H}_{\mathrm{im},l}\big{)}$.
This reconstruction is performed once to recover the object's real and imaginary phase, similar to~\cite{Tian2015a}.
Optimal values for $\tau_{\mathrm{re}}$,$\tau_{\mathrm{im}}$ were chosen based on manually evaluating the reconstructions from a range of regularization values.

\section{Results}
\label{sec:results}
\subsection{Reconstruction from Simulation}

\begin{figure}[t]
\centering
\includegraphics[width =1 \textwidth]{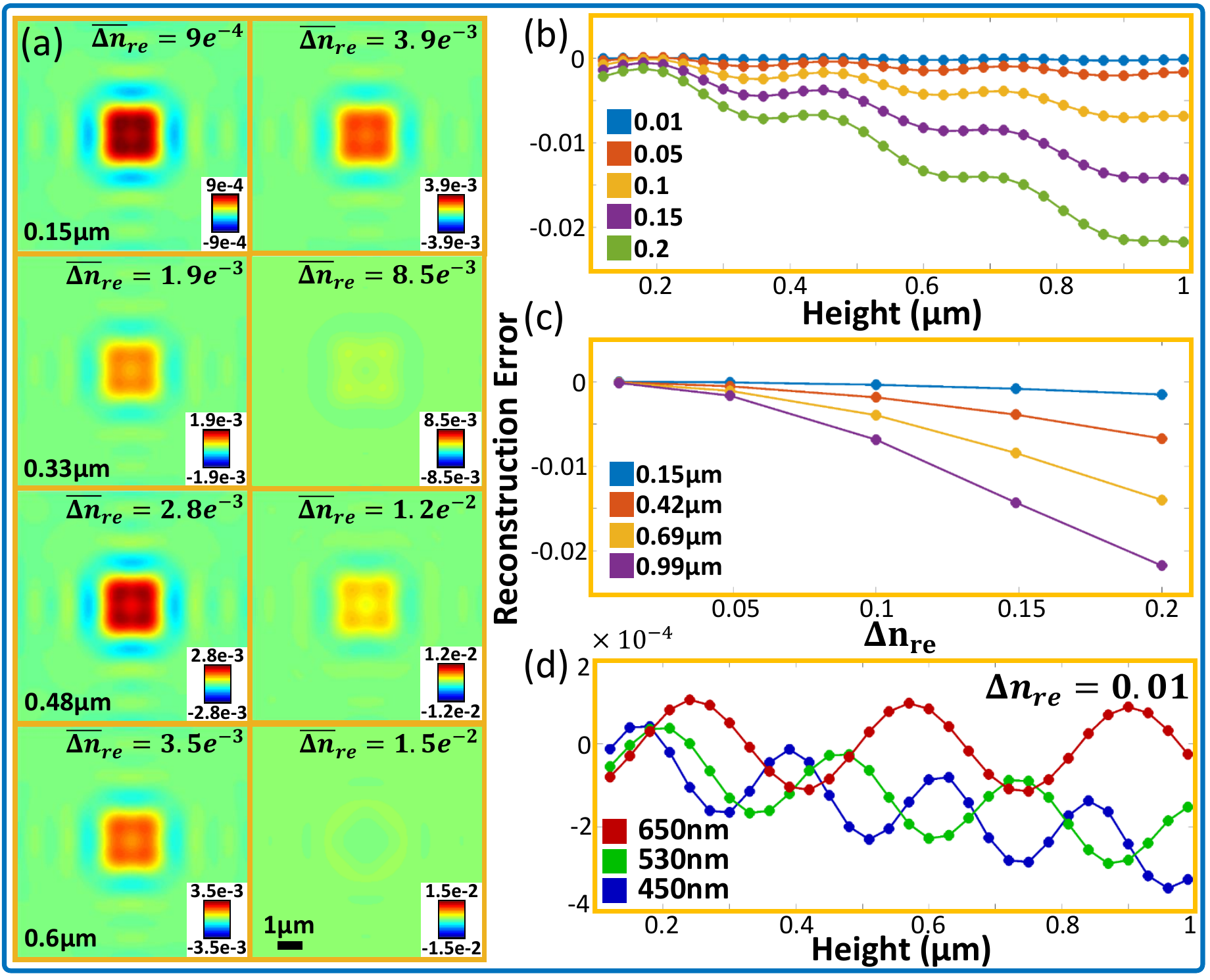}
\caption{(a) Linear model reconstructions of cuboid average permittivity contrast. 
		The color scales are adjusted based on the ground truth object's properties to show correctly recovered cuboids in red.
		 Weak permittivity (left) objects are more accurately recovered compared to strong permittivity (right) structures.
		 (b) Cuboid reconstruction error at $\lambda=530nm$ across different object heights at fixed permittivity contrast values. Nonlinear error is always present from backscattering, and increasingly tall objects quickly become underestimated from the enhanced sensitivity of our model.
		 (c) Cuboid reconstruction error at $\lambda=530nm$ across different permittivity contrasts for fixed object heights. The error is linear with permittivity contrast following Eq.~\eqref{UsProb}.
		 (d) Cuboid reconstruction error for a $\Delta n_{\mathrm{re}}=0.01$ object across multiple heights at different wavelengths ($\lambda = 450, 530, 650nm$). We observe nonlinear error also shifts with period following Fig.~\ref{F3}(d).}
\label{F4}
\end{figure}
We further evaluate our linear model's validity range by reconstructing the DDA-simulated objects.
We use the average pixel-wise difference between the reconstructed and ground-truth objects as our error metric.
Given the reconstruction's maximum bandwidth of 0.45NA, we filter the ground-truth object with a 0.45NA circular pupil to directly compare the best object resolvable with the system and the linear model's reconstructions.
In addition, we convert our recovered phase $\Phi$ and the ground-truth filtered object to the average RI contrast $\overline{\Delta n}$ for both real and imaginary components.
For an object of uniform permittivity contained entirely within the objective DOF, the average permittivity contrast is equivalent to
$\overline{\Delta \epsilon}(\bm{r}) = \Phi(\bm{r})/\mathrm{DOF}$,
where $\mathrm{DOF}=\lambda/\mathrm{NA}^2$. 
We subsequently convert this value to average RI contrast $\overline{\Delta n}$ based on the relation of permittivity and RI.
This choice enables direct comparison between our reflection model's reconstructions and the transmission reconstructions of Section~\ref{sec:results}. 
We convert our simulation outputs to the average RI contrast as well for consistency.
%

Visuals of the reconstructed cuboids are shown in Fig.~\ref{F4}(a) with color scales adjusted to match the expected average RI contrast for each object.
Fig.~\ref{F4}(b)-(d) shows the error for our reconstructions for fixed RI contrasts with object height (Fig.~\ref{F4}(b)), fixed heights with varying RI (Fig.~\ref{F4}(c)), and fixed RI contrast with object height across multiple wavelengths (Fig.~\ref{F4}(d)).
For weak RI contrast objects, we adequately predict the average of the object's phase and show primarily sinusoidal error due to the missing backscattering phase contribution (Fig.~\ref{F4}(a)-(b)).
With increasing RI values, the linear model underestimates taller objects as the forward-scattered field accumulates phase and exits the Taylor expansion's validity range (Fig.~\ref{F4}(a)-(b)).
With the missing backscattered phase from Eq.~\eqref{B}, we observe nonlinear, sinusoidal error as a function of object height with oscillation period varying with the illumination wavelength (Fig.~\ref{F4}(b),(d)).
The linearity with RI contrast is still preserved (Fig.~\ref{F4}(c)).

These results confirm our linear model's validity range is considerably more limited than transmission intensity-only approaches.
Without evaluating the object in an imaging medium of nearly equivalent RI, the reflection imaging case provides reasonable phase recovery only for thin structures at nanometer length-scales.
This is restrictive for biological sample imaging where structures, such as cells and bacteria, have varied size distributions that are both within and outside the validity range of this model.
This validity range is further reduced for objects with high variance in both size and RI, as the object's physical parameters would become ambiguous in our reconstruction.
Despite these factors, we show reliable object recovery on thin cellular structures with high contrast below.

\begin{figure}[h]
\centering
\includegraphics[width =1 \textwidth]{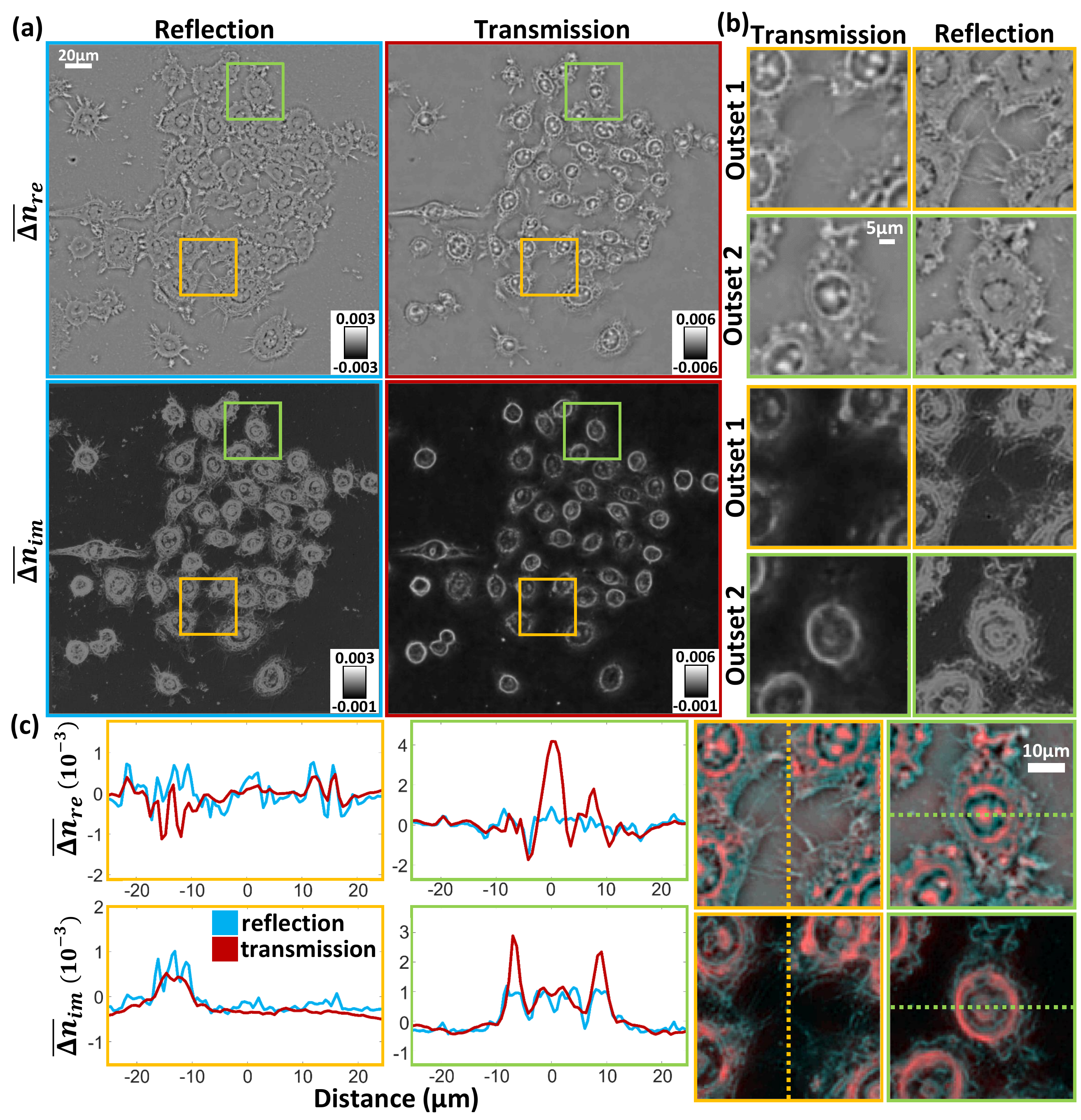}
\caption{(a) Full FOV complex RI contrast reconstructions for reflection and transmission (HeLa cells); 
			(b) Outset regions show (1)  cell boundaries and filopodia,  and (2)  cell nuclei reconstructions; 
			(c) Cross-sections compare reflection and transmission average RI contrast reconstructions and overlays of transmission and reflection reconstructions. 
			Outset 1 show better membrane structure contrast in reflection than transmission with agreement on the recovered average RI contrast values.
			Outset 2 shows reflection underestimates the RI contrast of tall nuclear features as expected from simulation.}
\label{F5}
\end{figure}

\subsection{Reconstructions from Experiment}
We experimentally investigate the validity range of our reflection model using the setup in Fig.~\ref{F1}(a).
A 530nm LED (Lighthouse LEDs, 10 Watt Jade Green) placed behind a white diffuser (Edmund Optics, 34473) and a 300$\mu$m pinhole (Thorlabs, P300H) composes the illumination source. 
This source resides inside a motorized (Thorlabs, Z806) XY translation stage (Thorlabs, ST1XY) and follows the K\"ohler geometry for oblique illumination up to 0.2NA.
This illumination angle was chosen based on the scannable light source's maximum range of motion.
A 4F setup provides $3.3\times$ magnification generating a 1mm diameter source (0.025NA, coherence parameter = 0.1) at the back-focal plane of the objective lens (10$\times$, 0.25NA, Nikon). 
The detection path collects the field through the objective and relays it to the camera (Thorlabs,CS2100M-USB) with a 200mm tube lens. 
The system's total NA is 0.45 with a $\sim$1.2$\mu$m lateral resolution. 
In each experiment, 97 images were acquired over an approximate five minute measurement period.
The samples consist of unstained HeLa cells fixed with ethanol (Fig.~\ref{F5}) and formalin (Fig.~\ref{F1}). 

For comparison, the transmission-based Intensity Diffraction Tomography (IDT) technique~\cite{ling2018high} was also applied to the same samples using 87 images and up to 0.25 NA illumination (630nm LED array) on the same objective lens. This system's specifications and reconstruction process are detailed in~\cite{ling2018high}.

We evaluate the recovered average RI contrast of the HeLa cells from both techniques in Fig.~\ref{F5}. 
Full field-of-view (FOV) reconstructions are shown in Fig.~\ref{F5}(a) with corresponding subcellular regions highlighted in outsets 1 and 2 (Fig.~\ref{F5}(b)).  
The cross-sections through these outsets are also presented in Fig.~\ref{F5}(c) with overlays directly comparing the two methods' reconstructions. 
Outsets 1 shows cellular membrane boundaries and filopodia and outset 2 highlights features within the cell nucleus.

Outset~1 highlights the improved contrast our reflection model provides over the transmission technique (Fig.~\ref{F5}(b)-(c)).
The observed filopodial structures range down to 100nm in diameter~\cite{porter1974scanning} while cellular membranes can be only a few nanometers in thickness. 
Comparing the reconstructions shows enhanced contrast from reflection and agreement in average RI contrast value with transmission for both the real and imaginary contrast (Fig.~\ref{F5}(c), orange).
These results indicate our reflection model reliably recovers these features despite the greater error from the height-dependent nonlinear back-scattered field.

Outset 2 shows our reflection model's underestimation of large objects.
Nuclear structures maintain similar RI contrast to the surrounding cellular material but are much taller than the surrounding membrane.
The green cross-sections in Fig.~\ref{F5}(c) show transmission measurements capture tall, higher contrast features while reflection underestimates the object RI in both the real and imaginary reconstructions.
This underestimation agrees with our results in Fig.~\ref{F4} where our linear reflection model could not adequately recover tall features at large RI contrast. 

The image overlays of Fig.~\ref{F5}(c) highlight the complementary nature of the transmission and reflection QPI measurements.
We observe dominant reflection reconstructions (blue) in the membrane features with weak phase while transmission (red) recovers nuclear features with larger phase.
Combining these modalities provides a more complete cell evaluation, where the cell boundaries become clearly visible from reflection and the nucleus is adequately measured in transmission.
This results suggests that this linear reflection model could still provide useful information for biological research applications when combined with transmission QPI modalities.

\section{Discussion and Conclusion}
\label{sec:discuss}
We derived and evaluated intensity-only linear scattering models for recovering phase in reflection from an object above a partially reflective boundary interface.
Our derivation showed both forward-scattered and back-scattered fields are measured in reflection from an illuminated object in the presence of this interface in similar fashion to the mirror tomographic approaches~\cite{guerin2004second,mudry2010mirror}.
These contributions provide slowly varying, approximately linear phase behavior in the forward case and rapidly oscillating, nonlinear behavior in the backscattering case dependent on object height.
Through rigorous DDA simulations, we confirmed this linear and height-dependent nonlinear phase behavior across objects of varying height, RI, and with different imaging wavelengths.
We presented a simplified linear model recovering only the {\it forward}-scattered phase from the object that accepts the nonlinear field behavior as error in the object reconstruction.
Through DDA simulations and object reconstructions, we showed the reflection case provides enhanced phase sensitivity and contrast for thin objects with weak permittivity contrast and underestimates taller, higher RI contrast structures.
We confirmed this result by measuring fixed HeLa cells on a glass slide in reflection and transmission using our approach and IDT, respectively.
This sample showed thin membrane structures are recovered with greater contrast in reflection, but tall nuclear features are underestimated with our approach when compared with transmission.
Our physical model shows linear approximations for intensity-only imaging systems are more restrictive in reflection than transmission due to the backscattered field's nonlinear behavior with object height.
Despite this limitation, very thin object features can be recovered with minimal reconstruction error using a linear model.
Given this model's strict limitations to thin objects, this intensity-only reflection phase system is best utilized in tandem with transmission modalities to provide thin feature recovery with the thick structures recovered in transmission.

The nonlinear behavior of the backscattered field can be better understood through an evaluation of the Ewald's sphere~\cite{streibl1984phase}.
The back-scattered field relates to the upper half of the Ewald's sphere containing high-frequency, high-resolution axial information about the object.
For low NA illumination in a quasi-monochromatic imaging system, this upper half of the sphere is effectively measured only at $\lambda/2$ with the back-scattered field.
This under-sampling results in oscillations with object height as observed in our DDA simulations.
Using higher NA objectives and illumination enhance this axial bandwidth to achieve improved axial resolution but requires additional hardware specialization and complexity~\cite{mudry2010mirror, simon2019versatile}.

Compared to the existing reflection interferometric modalities using temporal gating~\cite{yamauchi2008low, beaurepaire1998full, choi2018reflection}, our intensity-only system is limited by the sample thickness due to its lack of adequate depth sectioning as compared to the backscatter signal oscillation period.
Nevertheless,  performing phase imaging on thick samples using intensity-only measurements {\it without} temporal gating is still possible, as recently demonstrated by oblique backscattering microscopy (OBM)~\cite{Ford2012}. 
OBM exploits multiply scattered diffuse photons to generate oblique trans-illumination in a reflection imaging geometry~\cite{Ford2012}.
Consequently, OBM also relies on the dominant forward-scattering signal from the object, similar to our model.
The volumetric scattering medium effectively suppresses the backscattering contribution from the object, making it possible to establish a similar, yet less stringent linear model for quantitative phase recovery~\cite{ledwig2019epi,ledwig2018dual,laforest2019transscleral}.

One avenue for improving our model's validity range is through reducing the permittivity contrast between the object and surrounding medium.
Evaluating a cell sample in aqueous media with closer permittivity would solve this problem and enable researchers to evaluate their cultures directly, but a significant issue arises with the boundary interface.
With larger imaging medium permittivity, we reduce the mismatch between the medium and boundary layer that generates the forward-scattering necessary for this model.
The weakened forward-scattering field strength would reduce the linearity of the scattering field and also limit the model's validity range.
Using a higher permittivity boundary would be possible but necessitates specializing the imaging platform which makes it less accessible for biological research. 
In addition, the reflections from the aqueous media's surface would alter the physical model proposed here. 
For these reasons, we used the strongly scattering HeLa cell sample in our experiment to illustrate our technique's capabilities on glass slides commonly used for biological research.

Another avenue for improvement is to develop {\it nonlinear} scattering models that can better account for the scattering process.  
Computationally efficient and accurate multiple scattering models have recently been demonstrated for transmission-mode phase tomography~\cite{Tian.Waller2015,kamilov2015learning,Liu.etal2017c,tahir2019holographic,chowdhury2019high,lim2019high}. 
Adapting such models to reflection is possible~\cite{unger2019versatile} and will be considered in our future work. 

Beyond the {\it model}-based inverse scattering framework, the alternative {\it learning}-based tomographic reconstruction framework has drawn significant interest.  
Promising inverse scattering results have been demonstrated in transmission systems~\cite{li2018deep,sun2018efficient,li2018imaging,Goy201821378,rivenson2018phase,Xue_2019}.  A fruitful area of future research may be to apply these learning-based inverse scattering algorithms to reflection systems. 

\section*{Funding}
National Science Foundation (NSF) (1840990, 1846784), National Institute of Health (R01CA224911, R01NS108464, R21EY029412).

\section*{Acknowledgments}
The authors would like to thank Joy Muthami, David Kirk, and Napassorn Lerdsudwichai on the software design, testing, and validation of the physical imaging system. Alex Matlock acknowledges the National Science Foundation Graduate Research Fellowship (DGE-1840990).

\section*{Disclosures}
The authors declare that there are no conflicts of interest related to this article.
\bibliographystyle{abbrv}
\bibliography{Bib_combined}

\end{document}